\documentclass{egpubl}
\usepackage{eurovis2021}

\SpecialIssuePaper 

\usepackage[T1]{fontenc}
\usepackage{dfadobe}

\usepackage{cite}
\BibtexOrBiblatex
\electronicVersion
\PrintedOrElectronic
\ifpdf \usepackage[pdftex]{graphicx} \pdfcompresslevel=9
\else \usepackage[dvips]{graphicx} \fi

\usepackage{egweblnk}

\title[CommAID: Visual Analytics for Communication Analysis through Interactive Dynamics Modeling]{CommAID: Visual Analytics for Communication Analysis \\ through Interactive Dynamics Modeling}

\author[M. T. Fischer \textit{et al.}]
{\parbox{\textwidth}{\centering M. T. Fischer $^{1}$\orcid{0000-0001-8076-1376},
D. Seebacher$^{1}$\orcid{0000-0003-0097-5855},
R. Sevastjanova$^{1}$\orcid{0000-0002-2629-9579},
D. A. Keim$^{1}$\orcid{0000-0001-7966-9740},
and M. El-Assady$^{1}$\orcid{0000-0001-8526-2613}
}
\\
{\parbox{\textwidth}{\centering $^1$Universität Konstanz, Germany
 } 
}
}

\usepackage{todonotes}
\usepackage[autostyle=true]{csquotes}
\usepackage{xcolor}
\usepackage{wrapfig}
\usepackage{booktabs}
\usepackage{graphicx}
\usepackage{subcaption}
\usepackage{microtype}
\usepackage[inline]{enumitem}
\usepackage{soul}
\newcommand{\toolname}{\textsc{CommAID}}

\definecolor{ColorTeaserLabelA}{RGB}{46,84,150} 
\definecolor{ColorTeaserLabelB}{RGB}{197,90,17} 
\definecolor{ColorTeaserLabelC}{RGB}{83,129,52} 
\definecolor{ColorTeaserLabelD}{RGB}{248,55,92} 
\definecolor{ColorTeaserLabelE}{RGB}{230,180,40} 
\definecolor{ColorTeaserLabelF}{RGB}{255,128,0} 
\definecolor{ColorTeaserLabelG}{RGB}{136,0,255} 

\definecolor{ColorAppleGreen}{RGB}{155,220,71}
\definecolor{ColorAppleYellow}{RGB}{255,204,0}
\definecolor{ColorAppleTaleBlue}{RGB}{90,200,250}

\definecolor{ColorWorkflowRawData}{RGB}{130,130,130} 
\definecolor{ColorWorkflowVisualization}{RGB}{191,164,204} 
\definecolor{ColorWorkflowProperties}{RGB}{200,162,49} 
\definecolor{ColorWorkflowLevels}{RGB}{75,115,173} 
\definecolor{ColorWorkflowDomainExpert}{RGB}{70,70,70} 
\definecolor{ColorWorkflowProvenanceDAG}{RGB}{104,154,76} 
\definecolor{ColorWorkflowReport}{RGB}{130,130,130} 

\definecolor{distribution-level}{HTML}{FF8000} 
\definecolor{discussion-level}{HTML}{B5739D}

\DeclareRobustCommand\circledLetter[2]{
	\tikz[baseline=(char.base)]{
		\node[shape=circle,draw=none,fill=#1,text=white,inner sep=0.5pt] (char) {{#2}};
	}
}

\DeclareRobustCommand{\inlinesymbol}[1]{%
	\begingroup\normalfont
	\raisebox{-.2\height}{\includegraphics[height=1.5\fontcharht\font`\D]{figures/symbols/#1}}
	\endgroup
}

\usepackage[absolute,overlay]{textpos}

\begin{document}

\teaser{
 \includegraphics[width=\linewidth]{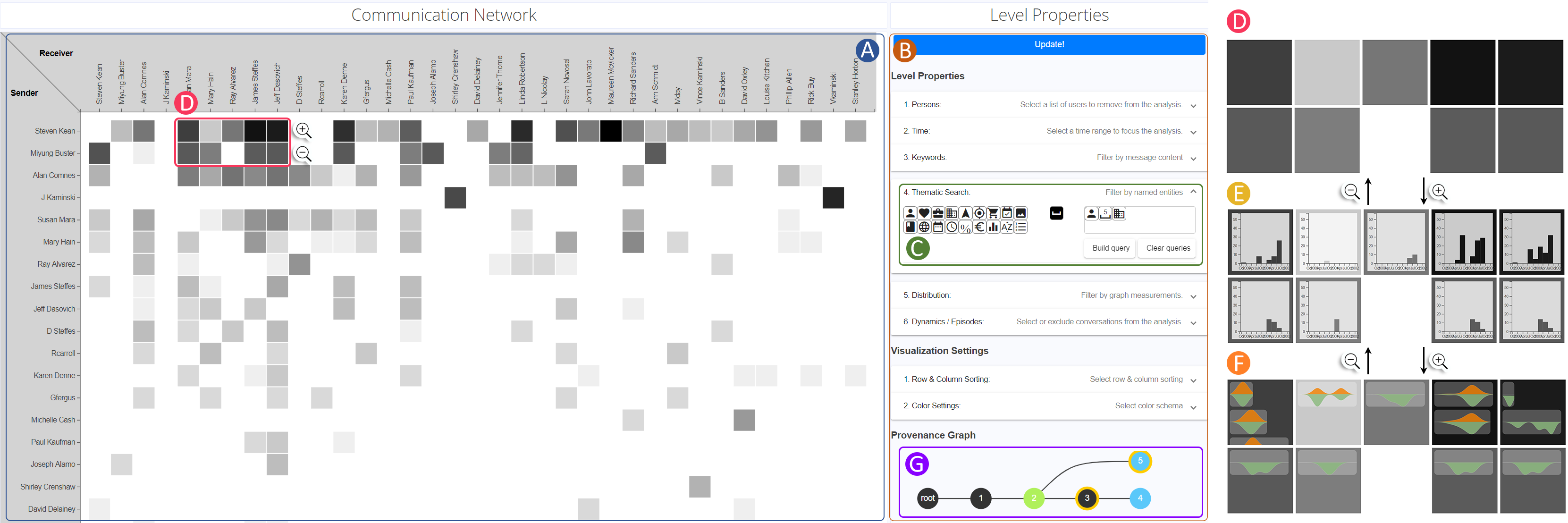}
 \centering
\caption{\toolname, an integrated visual analytics technique to analyze communication networks through dynamics modeling, semantic pattern retrieval, and an interactive multi-level matrix-based visualization\circledLetter{ColorTeaserLabelA}{A}.
	This visualizations enables the inspection of individual communication at different analysis levels through semantic zooming, while the linked level property pane\circledLetter{ColorTeaserLabelB}{B} allows to restricting the search space, using standard task filters, dynamics settings, and a thematic concept builder\circledLetter{ColorTeaserLabelC}{C}.
	The in-line views show the visualizations provided by the individual analysis levels (different zoom steps) presenting volume information\circledLetter{ColorTeaserLabelD}{D}, statistical distribution information\circledLetter{ColorTeaserLabelE}{E}as well as communication episodes using conversational dynamics\circledLetter{ColorTeaserLabelF}{F}.
	A provenance history graph\circledLetter{ColorTeaserLabelG}{G}allows to keep track of the analysis steps and results.
	The technique allows to interactively explore communication activity both from a metadata (connectivity, closeness, time, statistics) as well as a content level (keywords, thematic concepts) simultaneously, reducing discontinuities.
	\vspace*{6pt}
}
\label{fig:teaser}
}

\maketitle

\begin{textblock*}{17.5cm}(1.8cm,0.5cm)
	\tiny \noindent \emph{
	This is the \textbf{pre}-peer reviewed version of the following article: {Fischer, M. T., Seebacher, D., Sevastjanova, R., Keim, D. A., El-Assady, M. (2021). CommAID: Visual Analytics for Communication Analysis through Interactive Dynamics Modeling. Computer Graphics Forum, 40(3), 2021}, which has been published in final form at \url{https://onlinelibrary.wiley.com/doi/10.1111/cgf14286}. This article may be used for non-commercial purposes in accordance with Wiley Terms and Conditions for Use of Self-Archived Versions.}
\end{textblock*}

\begin{abstract}
 Communication consists of both meta-information as well as content.
 Currently, the automated analysis of such data often focuses either on the network aspects via social network analysis or on the content, utilizing methods from text-mining.
 However, the first category of approaches does not leverage the rich content information, while the latter ignores the conversation environment and the temporal evolution, as evident in the meta-information.
 In contradiction to communication research, which stresses the importance of a holistic approach, both aspects are rarely applied simultaneously, and consequently, their combination has not yet received enough attention in automated analysis systems.
 In this work, we aim to address this challenge by discussing the difficulties and design decisions of such a path as well as contribute CommAID, a blueprint for a holistic strategy to communication analysis. It features an integrated visual analytics design to analyze communication networks through dynamics modeling, semantic pattern retrieval, and a user-adaptable and problem-specific machine learning-based retrieval system.
 An interactive multi-level matrix-based visualization facilitates a focused analysis of both network and content using inline visuals supporting cross-checks and reducing context switches.
 We evaluate our approach in both a case study and through formative evaluation with eight law enforcement experts using a real-world communication corpus.
 Results show that our solution surpasses existing techniques in terms of integration level and applicability.
 With this contribution, we aim to pave the path for a more holistic approach to communication analysis.


\end{abstract}

\section{Introduction} \label{sec:introduction}
The enormous growth in the use of electronic devices and systems in the past decades has led to an exponential increase in digital forms of communication.
Simultaneously, the abundance of this digital communication~\cite{Scolari.DigitalCommunication.2009} and corresponding datasets has increased interest in how such communication can be analyzed in a wide variety of different domains, ranging from social sciences and digital humanities to engineering and business.
For example, it has been studied how social and psychological features change with computer-mediated communication~\cite{Gurak.PsychologyBlogging.2008}, how team performance can be assessed based on communication~\cite{Foltz.CommAnaTeams.2008}, how networks can be analyzed using text-mining~\cite{Yoon.TextMiningPatentNetwork.2004}, or how the evolution of dynamic communication networks can be visualized~\cite{Trier.DynVisCommNetworks.2008}.
This short list already shows a peculiar oddity when studying automated, digital communication analysis \emph{systems}: most existing approaches focus on either the content of communication \emph{or} on the network aspect---but not both.
The first group of approaches usually leverages methods from natural language processing~\cite{Manning.NLP.1999}, while the latter uses techniques from the field of Social Network Analysis~\cite{Scott.SocialNetworkAnalysis.2017}.

Interestingly, this distinction between content and network is neither present in the seminal works on human communication research~\cite{Moreno.HumanInterrelations.1934, WatzlawickBeavinJackson.Communication.1974}, nor in modern works~\cite{Pearson.HumanCommunication.2011} or current studies~\cite{Foltz.CommAnaTeams.2008, Mesch.SocialContextCommunicationChannels.2009}, even if digital methods aid the manual analysis.
Indeed, the analysis of network structure, communication patterns \emph{as well as} content plays an integral part~\cite{SavageDeutsch.TransactionFlows.1960, WatzlawickBeavinJackson.Communication.1974} of this research field.
Further, communication defined on a more abstract level can be described as the exchange of meaning between entities, transporting information encoded into symbols~\cite{Pearson.HumanCommunication.2011}, which refer to the content and its meaning while the overall context and transport modalities reflect the network and metadata aspect.
As such, analysis of the network/metadata or the content \emph{alone} can sometimes provide a limited, incomplete, or even biased view on the communication, which is not ideal.
Alternatively, to cover both aspects, the usage of two independent approaches would be required, which introduces domain discontinues and complicates search tasks, placing an additional burden on the users.

The problem of how both the network and the content perspective can be combined has not yet received enough attention when considering communication analysis systems.
This is especially relevant when such systems are used by non-communication experts, like in business intelligence applications or targeted criminal investigations, on which we based a case study (see Section~\ref{sec:evaluation_expert_assessment}) of our technique.
Typical tasks in these domains include searching for specific semantic content (e.g., negative product reviews, location names together with keywords), identifying groups (e.g., clusters or cliques), or a particular communication pattern (e.g., sequence, volume, timeline).
These tasks can be addressed with separate solutions.
But if, rather commonly, the search for specific semantic content needs to be restricted to a particular set of users (e.g., specific communication patterns, high centrality, or part of a specific clique), separate solutions struggle or even fail to support this use case.
Several other common tasks would benefit from such a combined search for cross matches and a fine-grained analysis of the communication network structure and context to detect communication behavior patterns and are therefore also not adequately supported so far.

\vfill\eject

In this work, we aim to address these shortcomings by discussing a possible technique as well as provide a framework for a holistic approach to interactive communication analysis.
We do not aim to describe a turnkey system or replace existing solutions, but rather discuss the challenges and design decision in such a system, present an exemplary blueprint prototype on how such an integrated system could look like, and gather expert feedback on such a broader approach, to support further research and positively influence system development in this domain.
For this, we extend upon two previous works, which we use as building blocks.
The first work~\cite{SeebacherFischer.ConversationalDynamics.2019} uses conversational dynamics to analyze communication patterns, covering the network analysis side.
The second work~\cite{Fischer.HyperMatrix.2020} describes a technique for hypergraph analysis, combining machine learning and a multi-level matrix-based visual interface.
From this, we borrow and adapt parts of the visual interface technique.

As part of this work, we present \toolname, making the following contributions:
\begin{itemize}
	\item A blueprint for a novel, interactive framework for a more holistic communication network analysis, providing a tight coupling between the network and the content analysis aspects, building upon individual models.
	To support a scalable and simultaneous exploration and analysis, we adopt a multi-level matrix-based approach with specific visualization levels and interaction techniques to seamlessly refine the search and support cross matches, with the support of a user-adaptable and problem-specific machine learning-based retrieval system.
	\item Describe two extendable models as example levels for network and content analysis: For the former, we extend our previous work on conversational dynamics~\cite{SeebacherFischer.ConversationalDynamics.2019}, while for the latter, we use established semantic concept detection from text mining.
	Additionally, several standard task levels are included.
	\item A discussion on the challenges, design choices, and future work for a holistic communication approach.
	\item One case study and an assessment with eight law enforcement experts using real-world communication data describing an application in the law enforcement field.
	The results demonstrate that our technique surpasses existing solutions, enabling the targeted and effective analysis of communication networks.
\end{itemize}

Our approach bridges the gap between network and content analysis in automated communication analysis systems, supporting domain experts in exploring and analyzing arbitrary bi-directional communication.
At the same time, we aim to pave the path for a more holistic approach to communication analysis.

\section{Related Work} \label{sec:relatedwork}
The \textbf{origins} of communication analysis can be traced back to the works of Simmel~\cite{Simmel.Soziologie.1908}, studying interaction in sociology, and Moreno~\cite{Moreno.HumanInterrelations.1934}, researching human networks and laying the foundation for social network analysis.
This technique, which describes a collection of research methods for identifying structures in systems, is widely described in the standard literature~\cite{Scott.SocialNetworkAnalysis.2017} and applied in many fields.
For \textbf{communication analysis}, the early works were later extended and taxonomized by Bavelas~\cite{Bavelas.CommPatterns.1950} and Leavitt~\cite{Leavitt.CommPatternsGroup.1951}, before Roger~\cite{Rogers.CommNetworks.1980} proposed to extend the field to communication networks.
While detailed, domains-specific content analysis, for example, in psychology~\cite{Berelson.ContentAnalyisComm.1952}, was already known almost seven decades ago; only the recent advancement of computational capabilities allowed the focus to shift to a bulk analysis of communication data on a larger scale.
Using methods from social network analysis, it became possible to investigate \textbf{network aspects} like social ties and communication behavior~\cite{Luo.SNACommChar.2015} by using centrality measures, detect communities~\cite{Xie.STAR.OverlappingCD.2013} and clusters~\cite{Aggarwal.Survey.ClusteringGraph.2010} or model whole artificial networks~\cite{Borgatti.NetwAnalysisSocialSciences.2009}.
However, using social network analysis on communication data primarily covers these network aspects.
When focusing on \textbf{metadata and communication content}, a virtually unlimited amount of analysis methods can be applied.
For example, metadata analysis~\cite{DeMontjoye.PrivacyBoundsMobility.2013} can be used to identify individuals, keyword-based searches~\cite{Yoon.TextMiningPatentNetwork.2004} can filter for specific content, while methods from natural language processing~\cite{Manning.NLP.1999} like sentiment analysis~\cite{Groh.SocialRelSentimentAnalysis.2011, Pang.OpinionSentiment.2008}, topic modeling~\cite{Rehurek.TopicModelling.2010}, or lexical chaining~\cite{Gold.ExplTextAnaLexicalEpisodes.2015} can support an advanced understanding of the meaning.

While this should give us amble scientific and technical methods at hand to analyze communication thoroughly, when we study automated, \textbf{digital human communication analysis systems}, we notice the peculiar oddity that
most existing approaches focus either on the network or the content aspect, but not both, as we discussed above.
A majority of the systems with communication analysis in mind follows the former approach.
The de-facto standard in civil security and business intelligence applications are IBM's i2 Analyst's Notebook~\cite{IBM.AnalystsNotebook}, and, to a lesser degree, the large network analyzer Pajek \cite{Batagelj.Pajek.1998}, both \textbf{commercial solutions}.
The open-source equivalent Gephi~\cite{Bastian.Gephi.2009} is also used sometimes.
While i2 Analyst's Notebook can be extended with content analysis capabilities, such a search is only offered as a separate interface.
From a \textbf{visualization perspective}, all follow a node-link-diagram based approach.
These suffer from inherent limitations like clutter or occlusion when the graph size becomes too large, and connections cannot be filtered enough for the search tasks, while techniques like edge bundling can only help so much.
In fact, a study by Ghoniem et al.~\cite{Ghoniem.ReadabilityNodeLinkMatrix.2005} shows that matrix-based visualizations are better suited for large or dense networks and perform better from a scalability viewpoint.
Various other methods are described in a general survey~\cite{Shiravi.SurveyVisNetwSecurity.2012} of visualization systems for large networks.
For example, when considering communication networks as multivariate graphs, one could employ techniques like Multilevel Matrices~\cite{vanHam.MultiLevelMatrices.2003}, Hyper-Matrix~\cite{Fischer.HyperMatrix.2020} or Responsive Matrix Cells~\cite{Horak.ResponsiveMatrixCells.2020} for improved scalability and detail, often combined with matrix reordering techniques~\cite{Behrisch.MatrixReordering.2016}.
Alternatively, it is possible to leverage semantic or magic lenses~\cite{Ghoniem.VAFLESemanticLense.2014} to highlight and enlarge relevant parts.
When focusing on social network exploration, the survey by Henry Riche et al.~\cite{HenryRiche.VisSocialNetwExpl.2010} focuses on specific extension for node-link and matrix-based approaches.

Looking at the \textbf{academic contributions}, we discover mostly alternative visualization and analysis methods.
\textbf{Matrix-like} approaches are used, for example, by GestaltMatrix~\cite{Brandes.AsymmRelSocNetw.2011} to visually analyze asymmetric relations, or MatrixWave~\cite{Zhao.MatrixWave.2015} for comparing multiple event sequences.
A notable set of approach that leverages matrix designs were proposed by Nathalie Henry:
MatrixExplorer ~\cite{Henry.MatrixExplorer.2006} presents the idea of combining node-link and matrix approaches, which NodeTrix~\cite{Henry.NodeTrix.2007} extends to address the occlusion problem for large node-link diagrams by switching to matrix view locally.
To address issues in path tracing in matrix views, they further present MatLink~\cite{Henry.MatLink.2007}.
\textbf{Timeline-based} designs were proposed as part of Timeline Edges~\cite{Reitz.RelationVisFramework.2010} to efficiently use edge space, in T-Cal~\cite{Fu.TCal.2018} to highlight areas with high communication volumes using distorted plot lines or as part of CloudLines~\cite{Krstajic.CloudLines.2011} to display event episodes in multiple time-series.
\textbf{Hybrid approaches} also exists, like Fu et al.~\cite{Fu.VisEmailNetw.2007} that propose to modify graph representations using multiple planes to recognizing communication patterns in e-mail networks.
When considering the metadata and content analysis side, countless methods exist in various fields.
However, many do not explicitly focus on communication analysis, and we will not discuss them here, although some can in principle be applied to a selection of the content-related tasks defined above (e.g., the interactive discourse analysis by Zhao et al.~\cite{Zhao.DiscourseAnalysis.2012}).

Leveraging analytical capabilities from both network and content information \textbf{simultaneously} has rarely been done.
Interestingly, \textbf{commercial systems} seem to be ahead of their academic counterparts.
Apart from Analyst's Notebook, which we discussed above, systems like Nuix Discover and Nuix Investigate~\cite{Nuix.DiscoverInvestigate.2020} for e-mail analysis and whole investigation frameworks like Palantir Gotham~\cite{Palantir.Gotham.2020} and more recently, DataWalk~\cite{DataWalk.2020} have become available.
Some have received mixed responses by the public given their primary application in the intelligence and law enforcement community.
As they are commercially developed, closed source solutions with few details about their detailed capabilities and internal workings, as well as their applications, they are often shrouded in secrecy (given their target domain).
This proves problematic because it hinders oversight from an independent community like academia to track capabilities or point out issues like bias, which becomes increasingly relevant with the usage of machine learning techniques within these solutions.
Looking at the \textbf{academic contributions}, we have three relevant approaches which combine network and content perspectives:
TimeMatrix~\cite{Yi.TimeMatrix.2010} by Yi et al. combines meta-information and network structure.
It uses a matrix-based visualization to analyze temporal social networks using TimeCells, showing a visual aggregate of a node's temporal information.
For example, it can show edge count for a pair of nodes over a period of time.
OpinionFlow~\cite{Wu.OpinionFlow.2014} by Wu et al. combines content with network structure analysis to visually analyze opinion diffusion.
They base their modeling on the network structure and sentiment-specific word embeddings, with the results using timeline-like visualizations.
IEFAF~\cite{Hadjidj.ForensicEMailFramework.2009} by Hadjidj et al. also combines content with network structure analysis.
It uses a multiple-coordinated view system with a node-link diagram as the primary visualization to support the forensic e-mail analysis, supporting various filter techniques, like metadata or keyword analysis and authorship attribution.

Given the little overlap between these solutions, their restricted applicability to communication in a generic case, and the growing support in commercial applications - in contrast to academic literature - reveals a missed opportunity.
This is the \textbf{gap} we aim to fill: Provide a blueprint for a more holistic approach to communication analysis that supports network, metadata as well as content aspects simultaneously by the use of extendable plug-in models in a single interactive visualization system, enabling the effective exploration of communication for interrelated tasks as defined above.

\section{Challenges and Design Decisions}
\label{sec:challenges_design_decisions}

Such an approach encounters several challenges.
One obstacle comes in the form of the different requirements to internal data representation and the \textbf{analysis methods}, like graph-based approaches or content-based methods as well as the communication type involved, and how they can be combined in a single system while acting on the same data set.
For this, we formalize communication modeling in Section~\ref{sec:modeling_communication} and describe the analysis as abstract operators working on a shared data space.
The second challenge concerns the \textbf{visual representation and interaction} when combining these different methods in a single framework.
The proposed system has to visually support different analysis modalities as part of a holistic framework through understandable and effective visualization methods, provide easy access to the visual results, and to allow useful interactions between them.
The visual design choice also depends on the size and sparseness of the communication network data.
For example, a design based primarily on \textbf{node-link diagrams} like IEFAF might work well for very small networks or larger networks when it is sparse and can be decomposed into interrelated communities.
However, choosing node-link diagrams makes it very hard to integrate additional information in an accessible way~\cite{Reitz.RelationVisFramework.2010}, with coloring, overlays, and details on demand as options.
An alternative, as proposed in the literature~\cite{Ghoniem.ReadabilityNodeLinkMatrix.2005}, is to use \textbf{matrix-based approaches} to support larger and denser networks, which also support in-cell content.
We will follow this path for our approach.
Compared to TimeMatrix, we extend the idea of using a matrix-based approach considerably by using multiple views involving semantic zooming within the matrix visualization and thereby displaying specialized visualizations in-line and on-demand.
Our approach scales well with the number of messages, which, as edges, are the primary source of clutter in a node-link-diagram.
However, when the number of users exceeds several hundred~\cite{Fischer.HyperMatrix.2020}, options like scrolling or magic lenses might be required.
As a general note, our design focuses on a holistic approach to communication analysis, compared to many existing approaches with limited interaction and exploration concepts or heavily adapted to a specific task.
The system is designed with flexibility in mind when analysis tasks require combining methods from different sub-fields of communication analysis.
When, however, analysis tasks are very specialized and, for example, are primarily related to exploring the network structure alone, a visualization based on node-link diagrams might be more suitable.
Further, a possibly viable alternative to our design choice would be to use \textbf{coordinated views}, providing spatially separate visualizations that are logically linked.
Such an approach could be explored further as part of future work (see Section~\ref{sec:discussion}).

\section{Modeling Communication}
\label{sec:modeling_communication}
In the following two sections, we describe the overall workflow of our approach, shown later in Figure~\ref{fig:workflow}.
We begin by defining requirements for an abstract analysis level (which we can think of as a model) in Section~\ref{sec:abstract_model} and define standard task levels to address common functionality.
We give two exemplary descriptions of more complex, extendable, and replaceable levels as a blueprint for individual communication analysis.
Firstly, a dynamics level in Section~\ref{sec:dynamics_level} to analyze network and metadata.
Secondly, a level for semantic concepts described in Section~\ref{sec:thematic_level}.

In Section~\ref{sec:investigation}, we then discuss the integration of the individual levels and the interactive exploration using visual analytic principles.

\subsection{Abstract Level}
\label{sec:abstract_model}
A communication network can be described as a multidigraph $G := (V, M)$, with $V$ a set of vertices representing the communication participants and $M$ a multiset of ordered pairs of vertices representing a communication event.
Additional metadata and content can be modeled by defining an information function $i: M \rightarrow D$, mapping a communication event to a data space $D$.
Individual analysis levels can now be generically defined as operators that act on the vertex space $V$, edge space $M$, and the information function $i$.
Each level can have none, one, or multiple in-line visualizations called views in the main interface (see Section~\ref{sec:system:visualization}).
These visualizations can transfer domain- and task-specific information relevant to a domain expert.
Further, each level has its parameters and filters for control.
As individual, separate levels itself would not provide many benefits, the key idea is to complement each other on the system scale.
Their flexible and simultaneous application in a single approach provides support for cross-matches, as level-specific filtering adds together to form a global filter.
Additionally, all levels can output a feature vector that is fed to a machine-learning-based retrieval system, described in Section~\ref{sec:ml_retrieval_system}, to enable intelligent user steering.
The system can be customized and extended to more specialized tasks by adding additional levels to cover more specific needs.

\subsection{Standard Task Levels \inlinesymbol{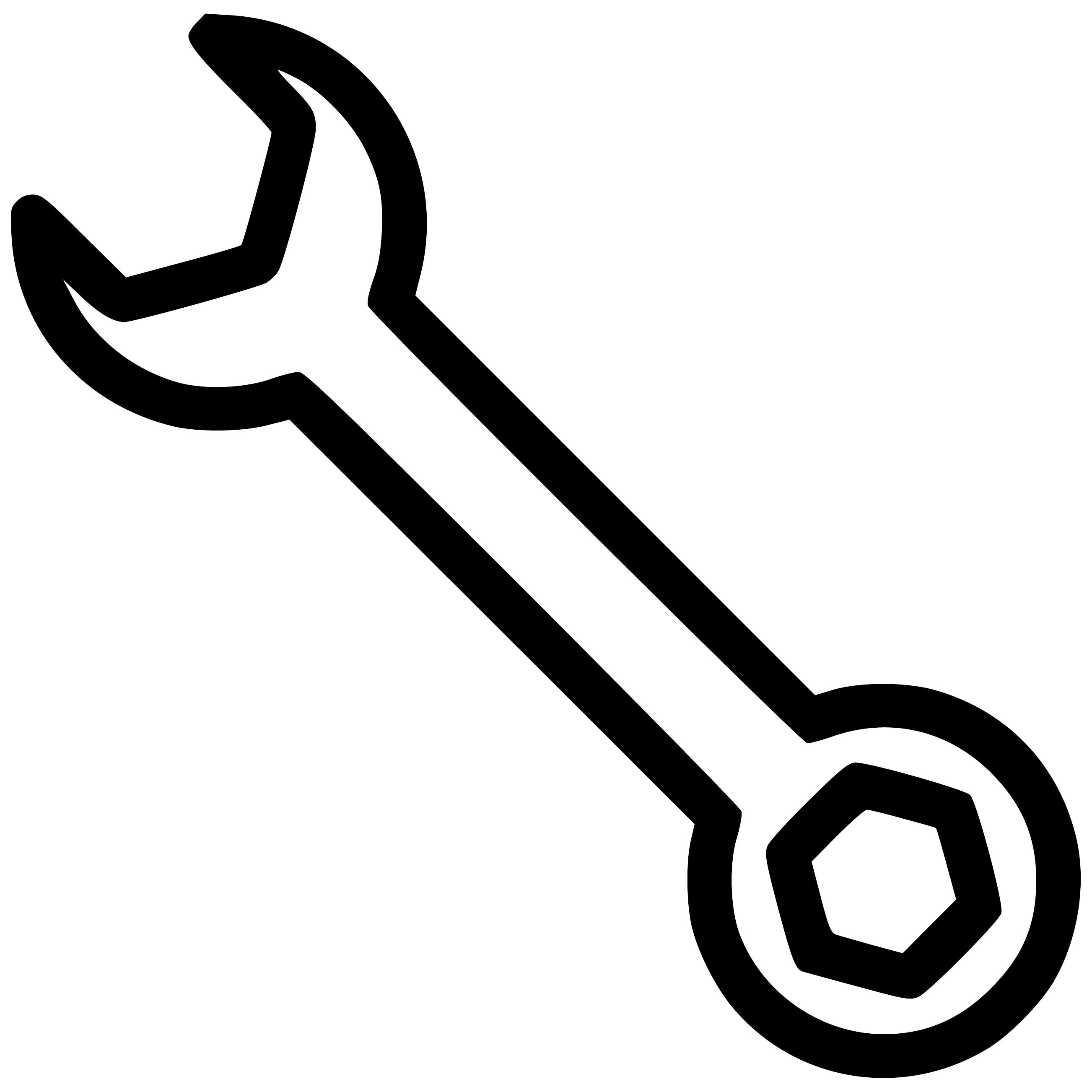}}
\label{sec:standard_task_levels}
Analysis usually requires a set of standard operators for filtering and selection, so we provide a set of standard task levels.
For example, to support simple tasks like restricting the time ranges, one can define an operator on $M$ which filters edges based on the timestamp information in the data space $D$ (a \textbf{timefilter level}).
Other examples are to filter participants in $V$ through properties in $M$ and $D$ (a user \textbf{selection level}), or keyword-based search by restricting based on content information in $D$ (a \textbf{keyword search level}).
As these levels act primarily as filters, a corresponding view (see Section~\ref{sec:system:visualization}) might not be required.
To provide basic visual analysis, on can define an operator on $V$ and $M$ (a \textbf{volume level}) which tracks the amount of communication between users, or an operator on $V$, $M$, and $D$, that track the temporal evolution of such communication (a \textbf{distribution level}).
For both, we provide corresponding views in the main interface (see Section~\ref{sec:system:visualization}).
For the remainder of this work, we will primarily focus on the more complex levels in the following two sections as they allow us to define task-specific analyses and only consider these standard task levels when necessary.

\subsection{Dynamics Level \inlinesymbol{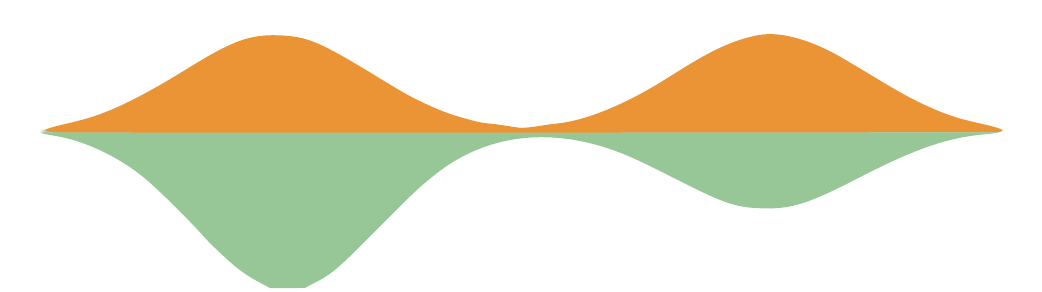}}
\label{sec:dynamics_level}
Different questions are of interest when analyzing the communication behavior between entities:
For example, how does the volume of communication develop? Is communication discontinued? Is it one-sided, or are there specific patterns?
However, if we look at communications only as individual messages, it may be difficult to answer such questions.
To analyze the dynamics of communication events more thoroughly, we follow our previous work~\cite{SeebacherFischer.ConversationalDynamics.2019}, and define a set of features which operate on the edge space $M$ and the information function $i$.
There, we model communication not only as individual events but as a flow, which can be described using distributions and a continuous density function.
This view enables us to easily model influence effects like the response time (both prior and delay), the width of the temporal influence, or control between spikes and general tendencies simply by adapting parameters like $\mu, \sigma$, or $h$, respectively. For details, see the original publication.
It further allows us to detect breaks in communication and thereby identify \emph{communication episodes}.
The choice for the granularity of the episodes is made globally depending on the dataset.
Together, this facilitates the structural communication analysis and helps to address the questions posed above.\newline
In our work, we leverage these previous ideas and adapt them to work on top of the abstract level formalization defined above.
We further add level properties and an in-line visualization (a view), which we further describe in Section~\ref{sec:system:visualization}.

\subsection{Thematic Level \inlinesymbol{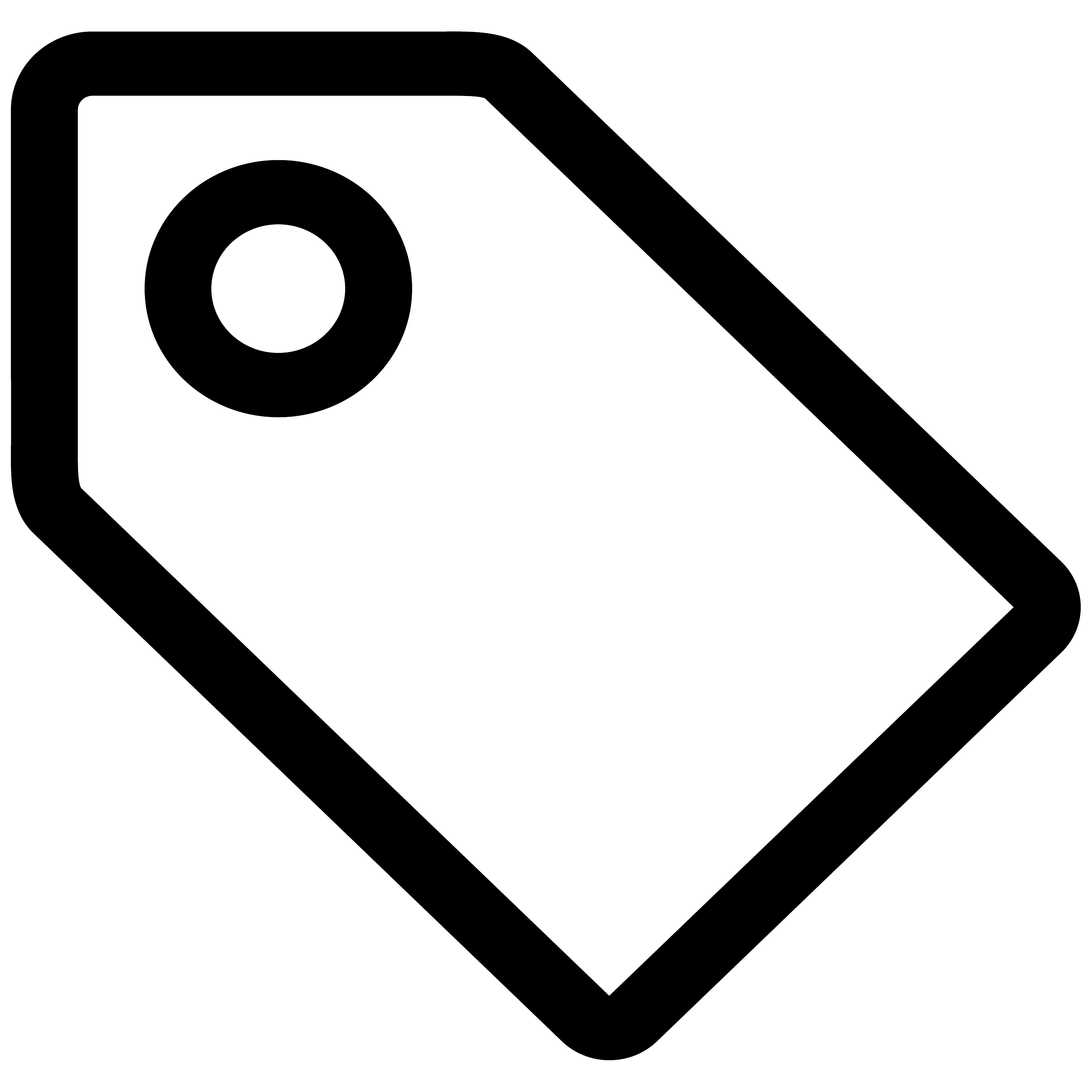}}
\label{sec:thematic_level}
In general, the inclusion of thematic concepts allows a user to refine their search task in a more powerful way than keyword lists and comes more naturally to analysts, who often think in concepts.
Regarding modeling, a thematic level operates on the edge space $M$, depending on the content information in the data space $D$.
A standard method to extract concepts from text data is named entity recognition.
Different languages can be supported by performing a language detection first and then chose an appropriate model.
There, it is possible to either use pre-trained models or adapt them with domain-specific or task-specific concepts.
However, a simple search using these semantic concepts might not be flexible enough to allow for more complex search tasks like "retrieve communication talking about a person in connection with a location".
Therefore, we propose an interactive visual query language that allows for a flexible combination of semantic concepts to fulfill such search tasks.
This query language allows creating multiple semantic queries based on spatial co-occurrence of semantic concepts.
These queries can then be combined using Boolean algebra to build more complex filters.
For example, the above search could be restricted further by additional requiring an organization to be mentioned, which is combined with the first query.
As this level acts more like a filter on the data, it is an example of a view-less level, not having a separate in-line visualization in Section~\ref{sec:system:visualization}.

\section{Visual Interactive Investigation of Communication} \label{sec:investigation}

\begin{figure}[b]
	\includegraphics[width=\linewidth]{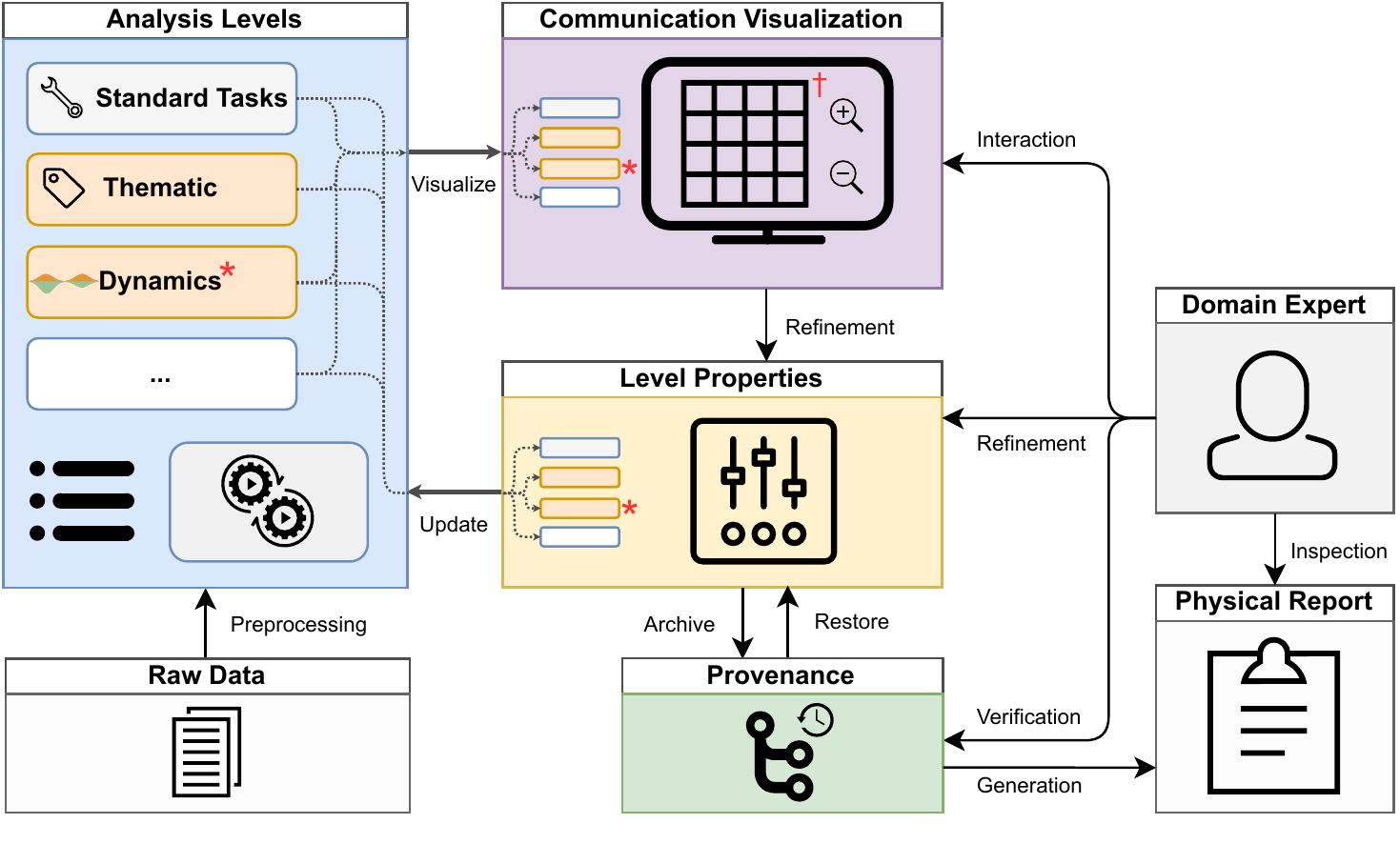}
	\caption{High-level workflow of our system, highlighting the main components and the interaction flow for the communication network analysis.
		The workflow begins with \textcolor{ColorWorkflowRawData}{\textbf{raw data}}~\inlinesymbol{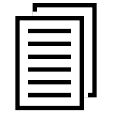}extraction and the generation of the individual \textcolor{ColorWorkflowLevels}{\textbf{level}}~\inlinesymbol{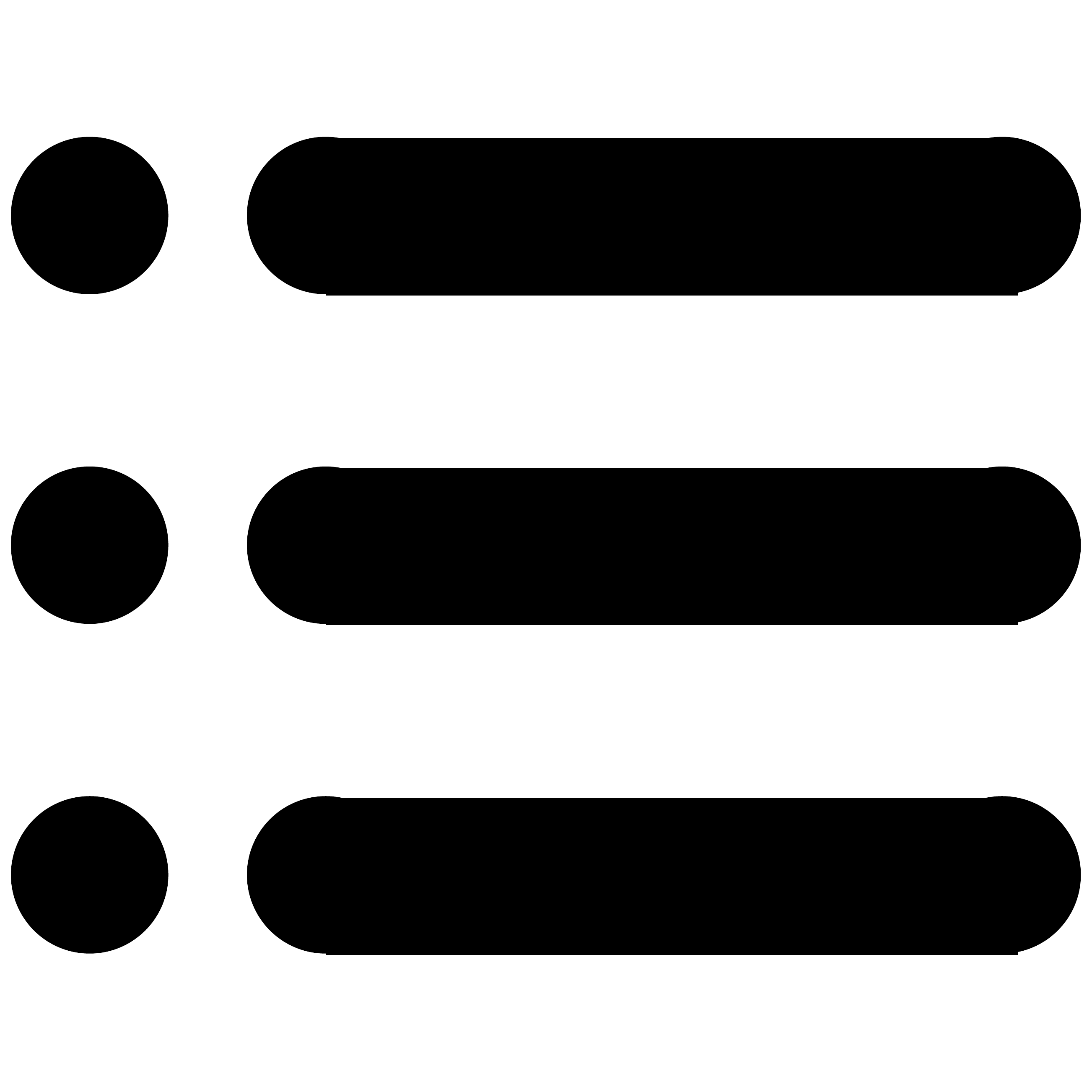}.
		A \textcolor{ColorWorkflowVisualization}{\textbf{multi-level matrix visualization}}~\inlinesymbol{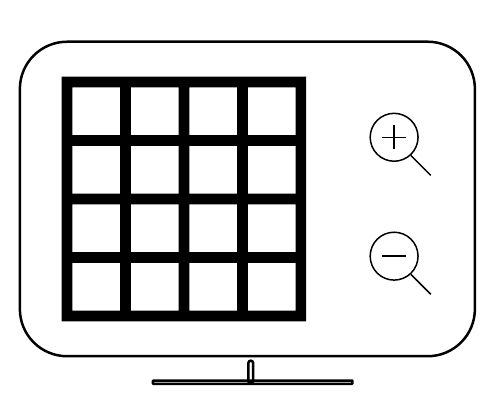}containing other in-line visualizations presents the current model state to the user and allows for different interaction and exploration schemes.
		The \textcolor{ColorWorkflowDomainExpert}{\textbf{domain expert}}~\inlinesymbol{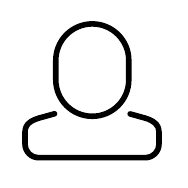}can then explore and refine the levels through their \textcolor{ColorWorkflowProperties}{\textbf{properties}}~\inlinesymbol{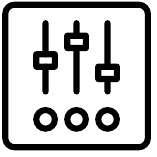}, leveraging an internal \textbf{relevance feedback system}~\inlinesymbol{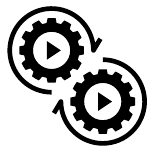}, updating the overall model state, and adapting the selection.
		The history of refinements is archived to provide \textcolor{ColorWorkflowProvenanceDAG}{\textbf{provenance information}}~\inlinesymbol{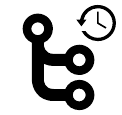}, which can be exported as part of a physical  \textcolor{ColorWorkflowReport}{\textbf{report}}~\inlinesymbol{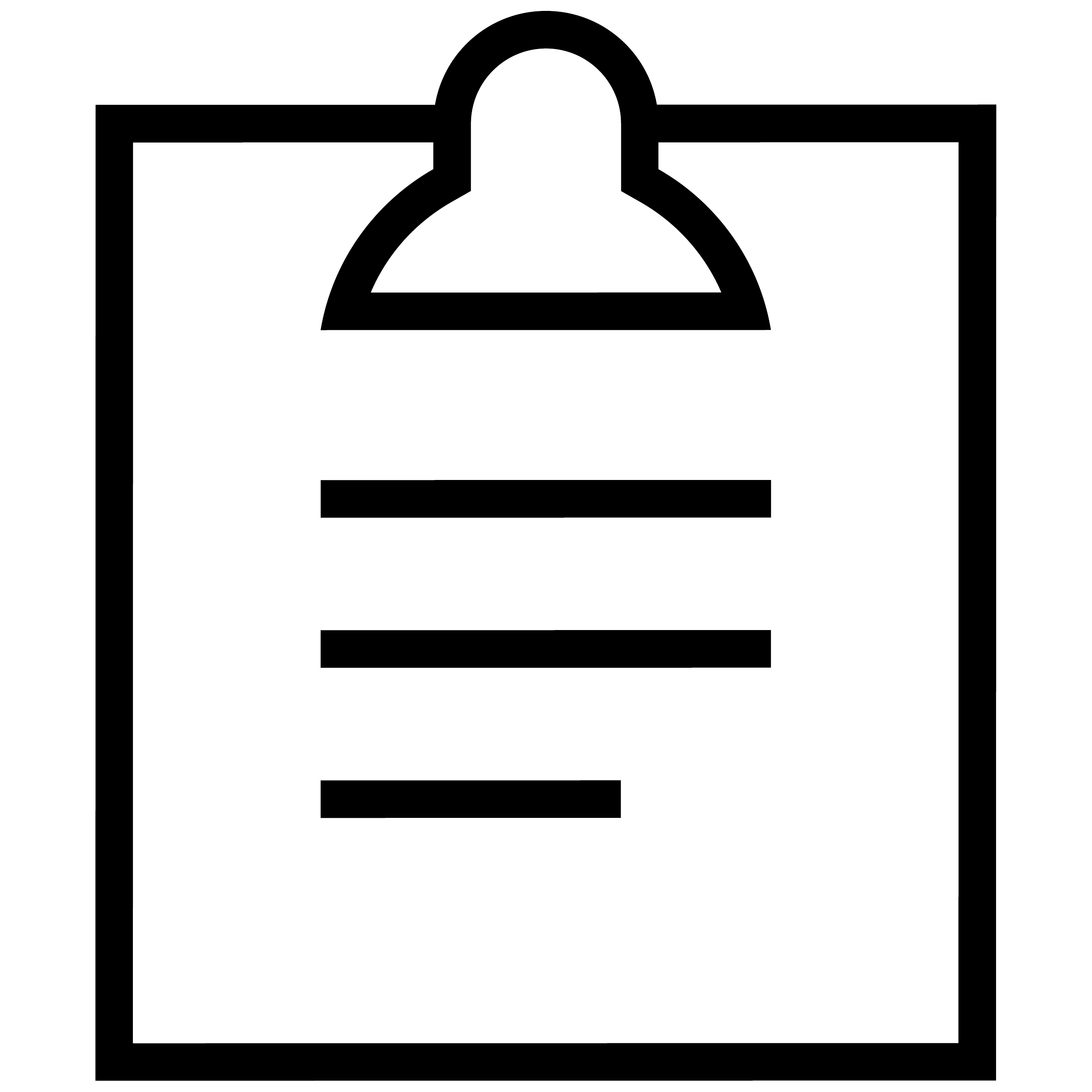}for inspection, traceability, and explainability.}
	\label{fig:workflow}
\end{figure}

This section focuses on the visualization and interaction concepts to integrate multiple levels in a single framework while providing a tight coupling between the network and content analysis aspects.
The proposed workflow for this architecture is described in Figure~\ref{fig:workflow}.
We begin by describing how the overall network can be visualized using a matrix-based visualization that provides multiple levels as views, representing the individual analysis levels' results.
Table~\ref{tab:components} shows the interplay between Levels, Views, and their properties.

\begin{table}
	\centering
	\begin{tabular}{l|c|c}
		\textbf{Level} & \textbf{View} & \textbf{Properties} \\
		\hline
		Volume & x & - \\
		Distribution & x & - \\
		Timefilter~\inlinesymbol{symbol_standard_task} & - & x \\
		User Selection~\inlinesymbol{symbol_standard_task} & - & x \\
		Keyword Search~\inlinesymbol{symbol_standard_task} & - & x \\
		\textbf{Thematic}~\inlinesymbol{symbol_thematic} & - & x \\
		\textbf{Dynamics}~\inlinesymbol{symbol_dynamics} & x & x
	\end{tabular}
	\medskip
	\caption{The different analysis levels in our system.
	Among the standard tasks levels, the two examples of more complex, custom level implementations are highlighted in bold.}
	\label{tab:components}
\end{table}

Conceptually, the information becomes more nuanced during level drill-down, going from overview to specific analysis to content, while each level addresses a specific question related to the level-modeling in Section~\ref{sec:modeling_communication}.
To facilitate the interactive exploration, the levels can be controlled via a property pane~\inlinesymbol{symbol_control}.
The levels can then act as filter methods, enabled through standard operators for standard task levels, steering options for conversational dynamics, and a visual query interface for thematic searches.
We specify how these individual levels can act together, which in-line visualizations they provide to support the exploratory analysis, and how user feedback for the adaptable retrieval system can support the search.
This methodology helps domain experts to gain a better understanding of the communication data by providing rapid-feedback through interactive filtering, covering different analysis levels simultaneously.
Finally, we describe how all steps are recorded in a provenance history graph, making the decision-making processes traceable.

\subsection{CommAID Interface Design \inlinesymbol{symbol_matrix_visualization}} \label{sec:system:visualization}
\begin{figure}
	\begin{subfigure}[t]{0.32\linewidth}
		\centering
		\includegraphics[width=\columnwidth]{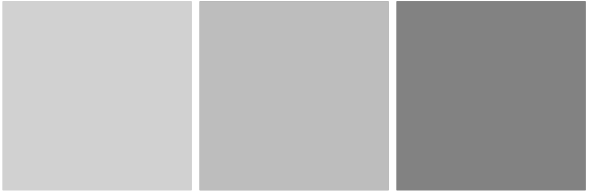}
		\caption{Volume View: The amount of communication is visually mapped to the color of the cell.}
		\label{fig:zoom-level-1}
	\end{subfigure}
	\hfill
	\begin{subfigure}[t]{0.32\linewidth}
		\centering
		\includegraphics[width=\columnwidth]{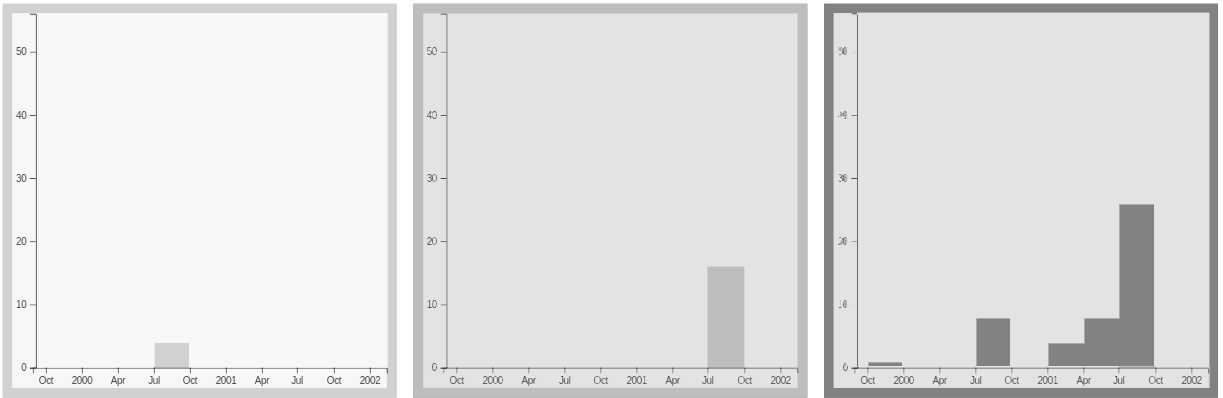}
		\caption{Distribution View: Provides an overview of the temporal distribution using barcharts.}
		\label{fig:zoom-level-2}
	\end{subfigure}
	\hfill
	\begin{subfigure}[t]{0.32\linewidth}
		\centering
		\includegraphics[width=\columnwidth]{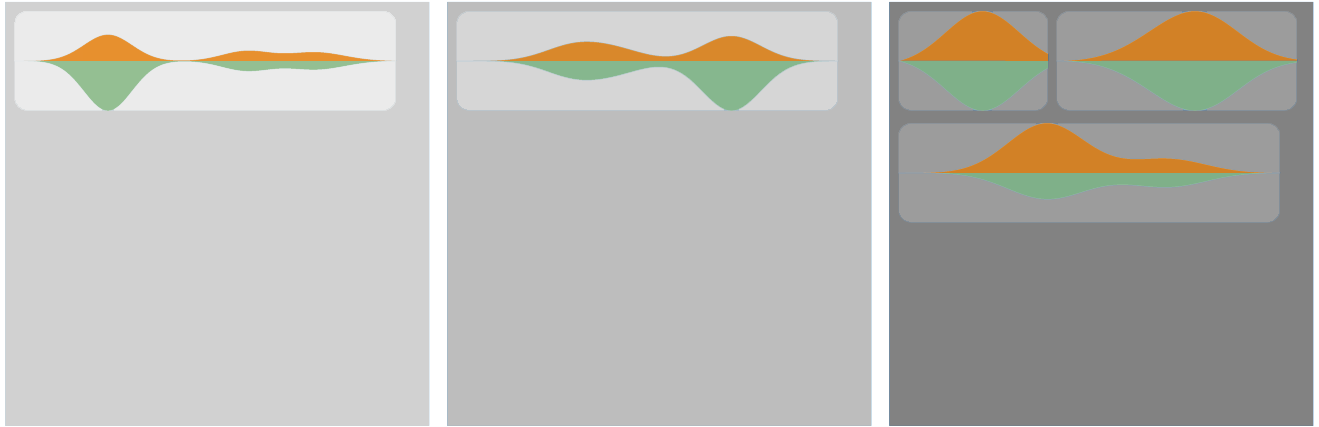}
		\caption{Dynamics View: Visualizes the communication episodes between two entities.}
		\label{fig:zoom-level-3}
	\end{subfigure}
	\caption{Overview of the three views we offer.
		We distinguish between the volume, distribution, and dynamics view.
		The first two in-line visualizations come from standard task levels (see Section~\ref{sec:standard_task_levels}), the latter from the dynamics level (Section~\ref{sec:dynamics_level}).}
	\label{fig:zoom-levels}
\end{figure}

For the visualization of the communication networks, we adapt our previous work~\cite{Fischer.HyperMatrix.2020} and use a multi-level matrix technique~\inlinesymbol{symbol_matrix_visualization}.
However, we change the meaning of the levels and the cell information: Instead of displaying increasingly detailed cell information, the in-line visualization represents the results of the individual levels discussed in Section~\ref{sec:modeling_communication} as independent views.
The overall interface of our approach is shown in Figure~\ref{fig:teaser}.
Apart from the interactive matrix-based visualization\circledLetter{ColorTeaserLabelA}{A}, the linked level property pane\circledLetter{ColorTeaserLabelB}{B} allows to restricting the search space, using standard task filters, dynamics settings, and a thematic concept builder\circledLetter{ColorTeaserLabelC}{C}.
In this prototype, three different views are provided: \textit{Volume}~\circledLetter{ColorTeaserLabelD}{D}, \textit{Distribution}~\circledLetter{ColorTeaserLabelE}{E}, and \textit{Dynamics}~\circledLetter{ColorTeaserLabelF}{F}.
They are shown as part of Figure~\ref{fig:teaser} and in more detail for three generic cells each in \autoref{fig:zoom-levels}.
A provenance history graph\circledLetter{ColorTeaserLabelG}{G}, discussed later in Section~\ref{sec:system:provenance}, allows to keep track of the analysis steps and results.
In the following, we explain the design rationale of the views.
It is important to note that through semantic zoom, the order in which the views are shown is fixed.

The basic principle of semantic zoom is that each cell of the matrix visualization~\inlinesymbol{symbol_matrix_visualization}serves as a canvas for a different type of analysis result of the communication between two entities in the network.
However, rendering detailed visualizations on the canvas of a cell only makes sense if a cell has a specific minimum window size.
Otherwise, even basic visualizations, such as a bar chart, are practically impossible to read.
Guidelines~\cite{Fuchs.DataGlyphs.2017} have been developed for the readability and required size for such visualizations.
Along those lines, and with the type of views in mind, we have chosen a view switch with every doubling of cell size.
When using a different type of view, the transition criteria might have to be adapted.
For example, either by using a different scaling factor or keeping the cell size for some view transitions and just switching the view.

\begin{figure}[!b]
	\begin{subfigure}[t]{0.48\linewidth}
		\centering
		\includegraphics[width=\columnwidth]{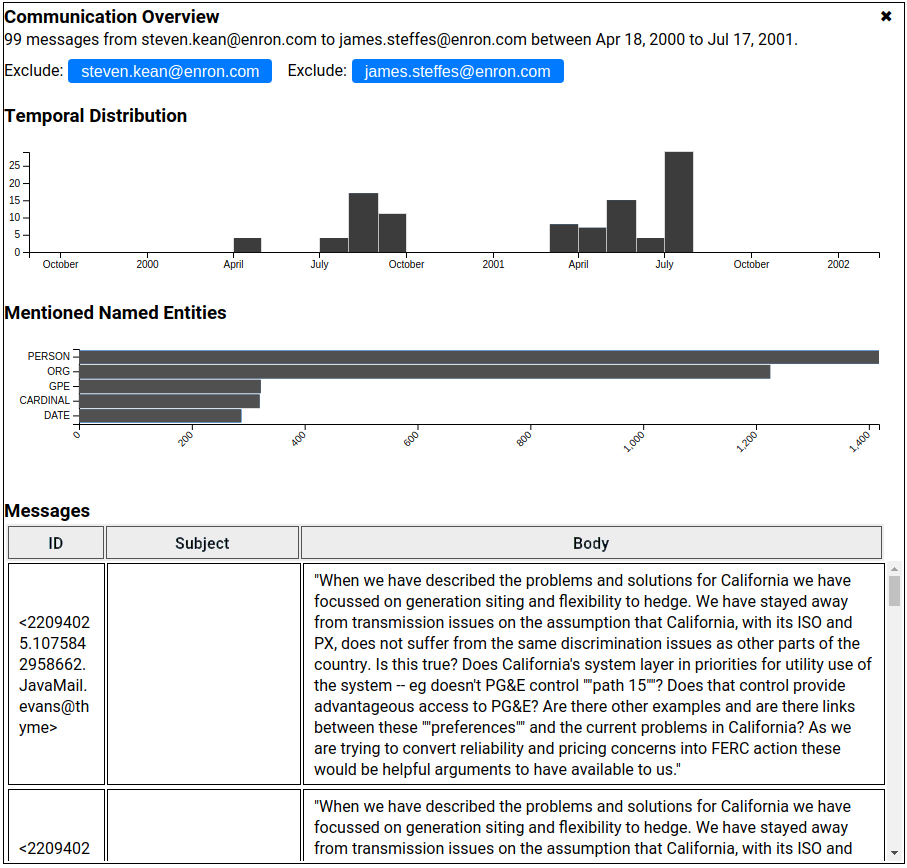}
		\caption{Details-on-demand populated by the distribution and thematic level presenting the time distribution, thematic named entities used, and, additionally, the raw data.}
		\label{fig:details-on-demand-1}
	\end{subfigure}
	\hfill
	\begin{subfigure}[t]{0.44\linewidth}
		\centering
		\includegraphics[width=\columnwidth]{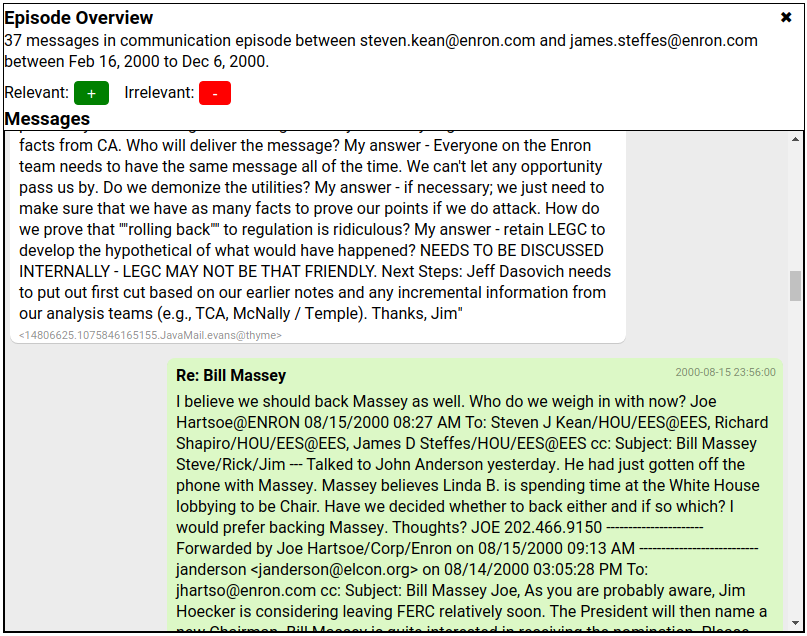}
		\caption{Details-on-demand for a communication episode visualizes the discussion in a chat-style metaphor. The communication can be ranked for the classifier discussed in Section~\ref{sec:ml_retrieval_system}.}
		\label{fig:details-on-demand-2}
	\end{subfigure}
	\caption{Overview of the details-on-demand offered by different semantic zoom levels, provided by different views.}
\end{figure}

The \textbf{Volume View} (belonging to the volume level) displays the number of communications between two entities, where the amount of communication is visually mapped to the cell's color.
Different color scales can be used depending on the task requirements.
\autoref{fig:zoom-level-1} shows a sequential, single-hue gray color scale, where white indicates that no communications are taking place, and black represents the maximum number of communications between two entities in the network.
Color schemes are replaceable, for example, for users with visual impairments or by using diverging color schemes to indicate deviations from the average.\newline
The \textbf{Distribution View} (belonging to the distribution level) is used to provide an overview of the temporal distribution of communications.
Similar to TimeMatrix, we use a bar chart.
Additionally, we use a background color to retain the information about the overall amount, matching the color used in the Volume View.
Thus, in addition to the temporal distribution of communications, global information can also be visualized.\newline
While these two represent views for the standard task levels, providing views for the custom models is especially interesting.
Here we offer the \textbf{Dynamics View}, visualizing results from the dynamics level~\inlinesymbol{symbol_dynamics}.
There, we represent the communication episodes between two entities in the network.
Depending on the tasks, the episodes can be shown chronologically or customly sorted.\newline
All three views have in common that they offer additional details-on-demand.
A click on a cell opens a zoom-level-dependent tooltip (see \autoref{fig:details-on-demand-1}), which provides information about the time distribution, named entities used, or raw data.
A click on an episode also opens a tooltip (see \autoref{fig:details-on-demand-2}), which visualizes the discussion content between two entities using a chat-style metaphor.
In both details-on-demand visualizations, the user can directly perform a refinement step, e.g., by excluding entities from the search or evaluating communication episodes for relevance.

\subsection{Level Properties \inlinesymbol{symbol_control}}

\begin{figure}[!b]
	\centering
	\begin{subfigure}[t]{0.475\linewidth}
		\centering
		\includegraphics[width=\linewidth]{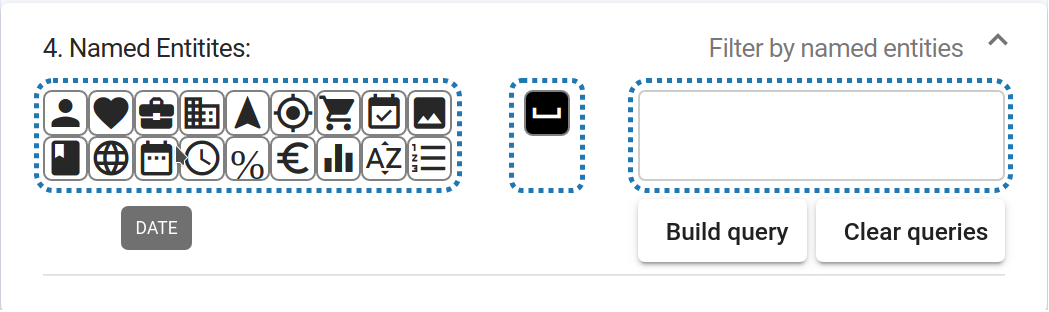}
		\caption{The named-entity repository as icons, a distance token and the query building component (f.l.t.r).}
		\label{fig:nlp-step-1}
	\end{subfigure}
	\hfill
	\begin{subfigure}[t]{0.475\linewidth}
		\centering 
		\includegraphics[width=\linewidth]{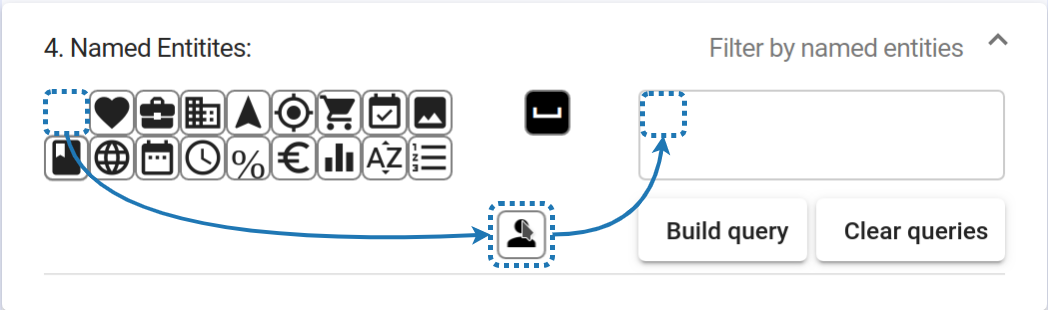}
		\caption{Building a named entity pattern query is performed via drag\&drop of icons and token.}
		\label{fig:nlp-step-2}
	\end{subfigure}
	\vskip\baselineskip
	\begin{subfigure}[t]{0.475\linewidth} 
		\centering 
		\includegraphics[width=\textwidth]{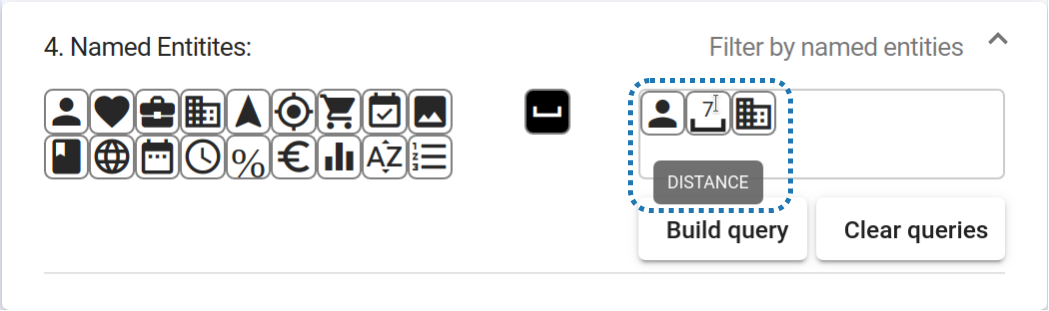}
		\caption{The concepts can be interactively rearranged, reflecting different query types. The maximum allowed word distance between two named entities can be set directly in the corresponding token} 
		\label{fig:nlp-step-3}
	\end{subfigure}
	\quad
	\begin{subfigure}[t]{0.475\linewidth} 
		\centering 
		\includegraphics[width=\linewidth]{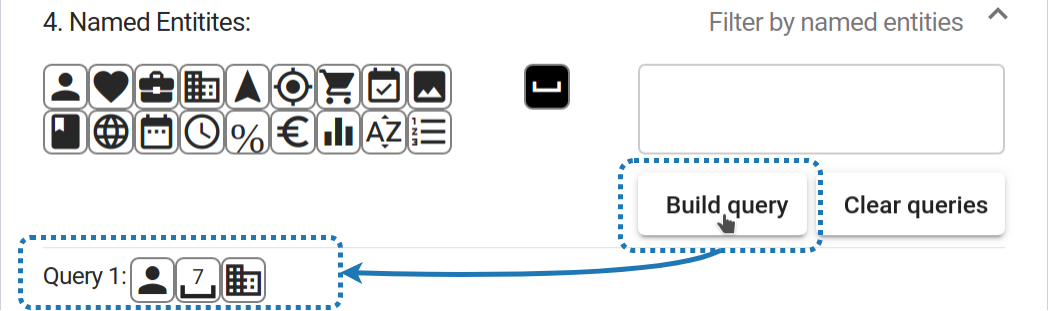}
		\caption{After all necessary corrections and adjustments have been made, the current query can be added to the set of applied queries. Using conjunctive (AND) or disjunctive (OR) combinations multiple queries can be combined.}
		\label{fig:nlp-step-4}
	\end{subfigure}
	\caption{Illustration for the step-by-step process for creating a named entity relation pattern query, searching for two concepts that occur within a given word length of each other.}
	\label{fig:nlp}
\end{figure}

Each level can have its own properties, accessible through a property pane~\inlinesymbol{symbol_control}on the right in the main user interface.
The standard task levels offer controls like cutoff values, include/exclude lists, or time sliders.
The dynamics level offers restrictions on the individual properties of conversational dynamics~\cite{SeebacherFischer.ConversationalDynamics.2019}.
Here, we want to describe one more complex property that can be offered on the property pane~\inlinesymbol{symbol_control}for custom levels, using the \textbf{Thematic Level} as an example: a visual query interface for thematic searches using named entities.
To generate the named entities, we employ a pre-trained model from spaCy~\cite{Spacy.NLP.2019}, containing a set of 18 named entity categories.
The interface itself is shown in \autoref{fig:nlp}, illustrating the individual components and the step-by-step process for creating a sample named entity relation pattern query, searching for two concepts that occur within a specified word distance.

As shown in \autoref{fig:nlp-step-1}, our visual query interface consists of three main components:
A repository containing a set named entities, such as persons, appointments, or organizations, a special token to allow distances between naming entities, and finally, a query building component.
To build a named entity pattern query, a user can drag one or multiple individual named entities from the repository and (optionally) the special token into the query building component, as highlighted in \autoref{fig:nlp-step-2}.
The concepts can also be rearranged inside the query building component, reflecting different query types, like single concepts, a chronology order of concepts, or distances between the concepts.
For the latter, the maximum allowed word distance between two named entities can be set directly in the corresponding token in the query builder.
In the example shown in \autoref{fig:nlp-step-3}, the maximum allowed distance between named entities is set to seven words.
After all necessary corrections and adjustments have been made, the current query can be added to the set of applied queries (\autoref{fig:nlp-step-4}).
The query shown only serves as a single example of a named entity query, where other types are possible.

\subsection{Machine-Learning-based Retrieval System \inlinesymbol{symbol_gears}}
\label{sec:ml_retrieval_system}
The design of using multiple complementing levels allows for cross-level search, with level-specific properties that can act as filters, acting together to form a global filter.
This already allows for more powerful and interrelated search tasks than the application of individual levels alone.
However, depending on the types of filtering defined and their interactions, the possible settings might overwhelm domain experts.
Therefore, we propose that each level can output a feature vector for a communication event.
Level-specific vectors can be combined to a single, large feature vector used for classification purposes in a user-steerable machine-learning level.
Although progress has been made to use deep learning efficiently by reducing training time~\cite{Salimans.WeightNormalizationDeepLearning.2016} and improve explainability~\cite{Guidotti.SurveyBlackBoxModels.2018}, their usage is still problematic when requiring (theoretical) traceability, for example, due to legal constraints.
Consequently, as proposed in the literature~\cite{Ming.RuleMatrix.2019}, we employ a rule-based approach based on a random forest model.
However, while this can fulfill legal requirements, from a perspective of lay use, a random forest's decisions might still be tough to understand.
It has, however, the additional benefit that the training size can be relatively small (usually much less than a few dozens), make it suitable for an interactive application, while the examples can easily be collected by the users themselves, based on their expert knowledge.
A user can train individual automatic classifiers that support him on specific tasks and modularly combine them to overall predictions.

\begin{figure}
	\begin{subfigure}[t]{0.31\linewidth}
		\centering
		\includegraphics[trim={0 13cm 0 0},clip,width=\columnwidth]{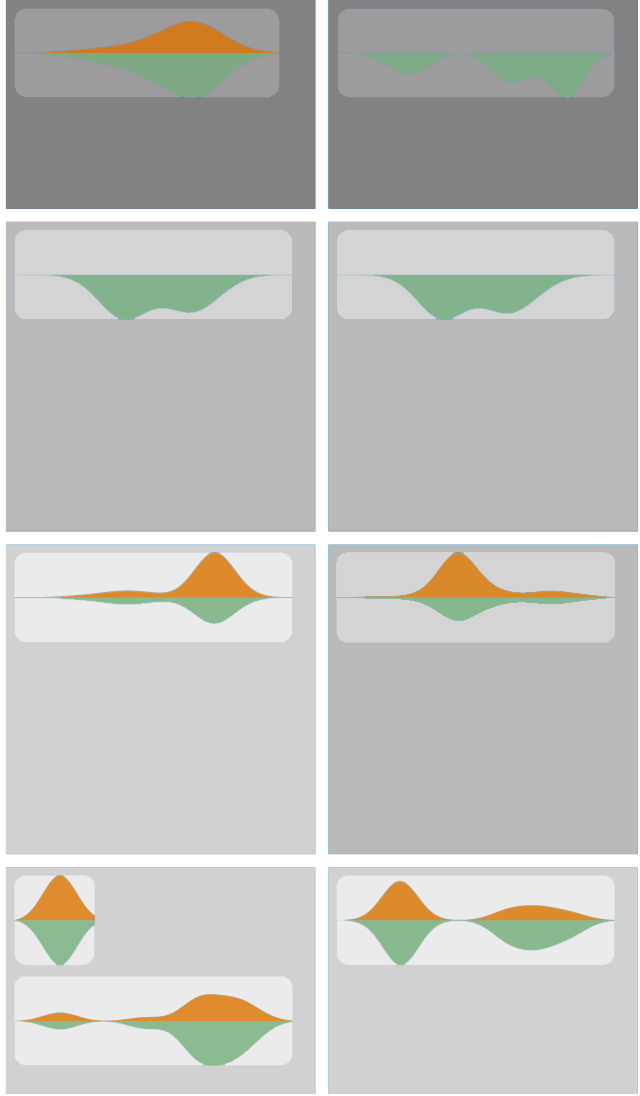}
		\caption{Cutout showing the individual episodes.}
		\label{fig:ml-1}
	\end{subfigure}
	\hfill
	\begin{subfigure}[t]{0.31\linewidth}
		\centering
		\includegraphics[width=\columnwidth]{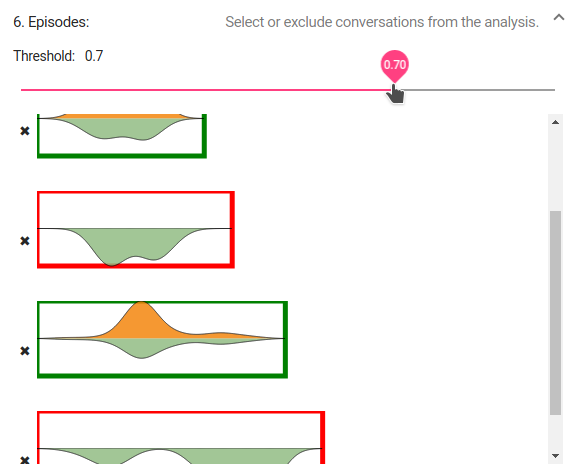}
		\caption{The positive and negative samples.}
		\label{fig:ml-2}
	\end{subfigure}
	\hfill
	\begin{subfigure}[t]{0.31\linewidth}
		\centering
		\includegraphics[trim={0 12cm 0 0},clip,width=\columnwidth]{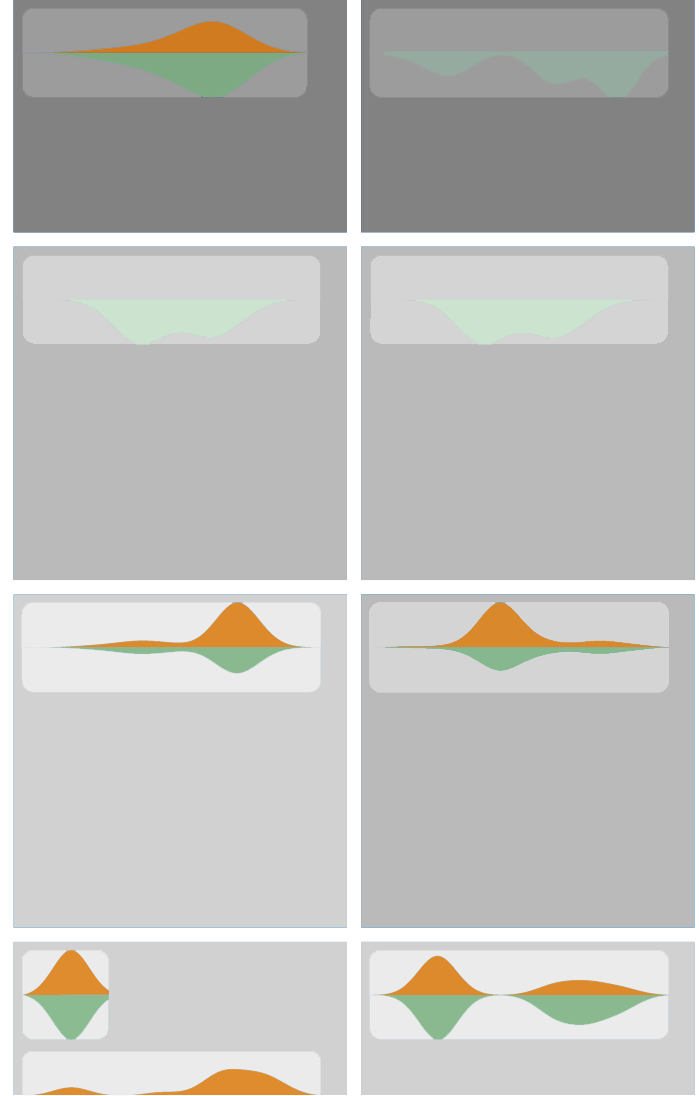}
		\caption{Results, with irrelevant episodes faded.}
		\label{fig:ml-3}
	\end{subfigure}
	\caption{By providing feedback, users train ML models to identify relevant conversational dynamics in episodes. Here, the aim is to identify episodes in which the selected groups start the conversation, leading to a discussion of both entities.}
	\label{fig:machine-learning}
\end{figure}

The selection of training examples for the classifier happens interactively.
A user can label communication binary as relevant or irrelevant to perform a semi-automatic classification of communication into user-defined classes.
An example of such a selection is shown in \autoref{fig:machine-learning}.
Such a trained classifier can then perform the binary classification for all other communications, acting as an additional high-level filter.
Since we use a Random Forest Classifier, we can model the uncertainty for the prediction, which is useful for threshold.
This also allows presenting ambiguous communication, where the classifier is very uncertain, to the user for re-labeling, allowing for an interactive optimization.
To separate between this semi-automatic retrieval system and manual level property settings, communication that is filtered out by the automatic system is only faded out based on a variable threshold, but not hidden completely.

\subsection{Analytical Provenance \inlinesymbol{symbol_provenance_dag}} \label{sec:system:provenance}

\begin{figure}[t]
	\centering
	\includegraphics[width=\linewidth]{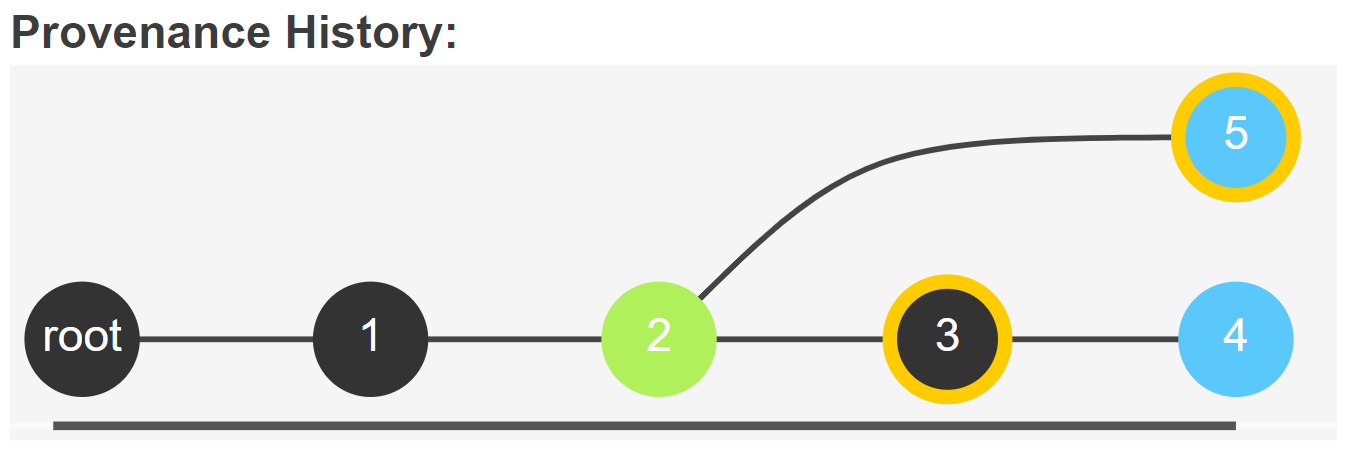}
	\caption{Provenance history component showing previous filter states as nodes in a directed acyclic graph (DAG).
		The {\color{ColorAppleGreen} currently selected node} and the {\color{ColorAppleTaleBlue} leaf nodes} are specially colored and important states can be {\color{ColorAppleYellow} starred} for easier distinction.
		The states allow for more flexible navigation, to revert from dead ends or branch of to a different analysis direction.}
	\label{fig:provenance_history}
\end{figure}

To support experts even in lengthy and complicated investigations, we offer a provenance history component, which is displayed in Figure~\ref{fig:provenance_history}.
Since explainability is relevant, for example, in court cases, decision making and a record of the obtained results must be preserved.
The provenance history contains previous filter states as nodes in a directed acyclic graph (DAG) to allow for more flexible navigation compared to a linear history~\cite{Haeussler.UrbanTrafficSensor.2018}.
Important states can be visually starred.
The user can navigate between different states, go back to previous results, or branch off as a new starting point for further analyses.
This enables the user to continuously verify and retrace results, which is especially advantageous for creating trust in the user's results.
Further, the individual filter states can be bundled into a physical report and thus archived.
In this way, the analyst's results can be reproduced, retraced, and explained, even after an extended period and/or to third parties.
Since each of the steps and all of its belonging information and metadata and the obtained results can be reviewed and analyzed independently, this allows for explainability of the results obtained.

\section{Domain Expert Assessment}
\label{sec:expert_interviews}
To demonstrate the effectiveness and improvements compared to existing approaches of the visual exploration of communication behavior in \toolname{}, we conduct an expert assessment of the prototype while additionally conducting a case study together with a small-scale user study later on.
As communication data, we use the largest publicly available source, the Enron dataset~\cite{Dataset.Enron.2004}, encompassing 517\,431 messages from 151 users.

\subsection{Formative Expert Assessment}
\label{sec:evaluation_expert_assessment}
The assessments were conducted by demonstrating the prototype to six domain experts (LEA~1--3, RS~1, SI~1--2).

\textbf{Expertise}
LEA~1 is a criminal investigator at a European law enforcement agency with extensive experience in the field, including communication and network analysis with graph-based visualization using commercial systems like IBM i2 Analyst's Notebook, Pajek, and Gephi.
LEA~2 is also a criminal investigator and has no prior experience using graph-based visualizations for communication analysis.
He focuses more on the communication content, using forensic tools like Cellebrite and IBM i2 Analyst's Notebook, which is laborsome.
LEA~3 is a senior judicial commissioner in law enforcement with extensive experience in digital investigation techniques.
He is aware of the systems and graph-based approaches used within his unit but has limited experience using them himself.
RS~1 is the head of a university-affiliated institute for policy and security research, a full professor and senior researcher working on government projects.
He has worked with graph-based visualizations and approaches for over 20 years, for instance, for bibliometric investigations.
SI~1 is a senior project lead within the security industry for developing investigative solutions for LEAs.
SI~2 is a junior solutions specialist within the security industry and worked on visualization techniques, including graph-based visualizations, for criminal investigations before.
All but one (SI~2) of the experts each have more than 15 years of experience in digital and criminal investigations.

\textbf{Methodology}
The expert assessment was conducted as a formative evaluation taking 90 minutes, with the experts observing, commenting, and asking questions on the system, while provided with an user sheet.
They were also allowed to steer the exploration by requesting specific actions.
As such, a formative evaluation without direct usage usually cannot replace the benefits of a full user study; we, therefore, additionally conducted a user study with two further domain experts later on, as described in Section~\ref{sec:evaluation_user_study}.
The experts were first given a ten-minute introduction about the aim, which is aligned with the tasks defined in Section~\ref{sec:introduction}:
facilitating the visual analysis and exploration of large amounts of communication data using different visualization and filtering methodologies simultaneously to structure and reduce the search space to a manageable size for inspection by an analyst.
The prototype was then showcased and explained during 30 minutes of interactive demonstrations, where the experts actively asked questions and commented on the presented aspects.
Presented aspects include using the network overview to detect promising connections and explore individual details and communication episodes in level-specific visualization through drill-down, which we then discussed with the experts.
Further, we presented and then debated the different interaction techniques and the available filters that can be used to reduce the selection.
Finally, the example-based machine learning retrieval classifier was discussed.
The interactive session was followed by a structured interview (see appendix), taking about 50 minutes, using a set of 29 prepared questions about various aspects of the approach.
This interview was interluded by interactively presenting aspects in the demonstrator when requested by the experts.
The whole session aimed to elicit experts' opinions on our system's design and interaction decisions and identify aspects that they find helpful or prone to misinterpretation.
We were also interested in how they would apply these methods in their specific workflows and criminal investigations in general.
The findings of these assessments and their comments are described in the following.

\textbf{Findings}
All the experts state that both the approach of using a \textbf{matrix-based overview visualization} and using a semantic zoom for more details is a new approach in their domain.
For example, according to LEA~1, he has \enquote{always worked with graph tools} so far and thinks of our technique as \enquote{very interesting and helpful}.
Together with the other experts, he thinks that a matrix-based visualization is superior to graph-based approaches in \enquote{terms of scalability} (SI~2) and displaying \enquote{supporting information} (RS~1).
Improvements recommended by LEA~1 and LEA~2 are that the matrix columns are freely reorderable.
Regarding the \textbf{semantic zooming}, the experts are familiar with such a concept from everyday applications like digital maps.
For communication analysis some did not expect this functionality at first (cf. LEA~2).
However, it supports their work and is an excellent way to drill down to \enquote{go into the raw data} (cf. LEA~2).
The semantic zoom was judged intuitive (cf. RS~1) but partly expected, based on his decade long experience with networks, that - instead of the communication structure and content - more information about the \enquote{importance of relations} (RS~1) is shown.
RS~1 proposes to include such information as another level, as the design is \enquote{flexible enough} (RS~1).
This could indeed be realized by adding more than our two example \textbf{analysis levels}, for example, an additional level for centrality analysis.

\vfill\eject

In terms of \textbf{filtering}, the experts are happy to have the standard task functionality included.
However, more advanced concepts like the semantic named entity search were \enquote{unexpected} (LEA~1, LEA~3).
In their previous experience, the LEA's were only able to search using lists of keywords and were \enquote{never able to search for concepts} (LEA~1).
They regard this functionality to have much potential, as it allows for \enquote{more generic} (LEA~1) and high-level search terms.
As the prototype's current implementation restricts them to AND-based queries, all experts state they would like to combine the queries more freely using Boolean logic.
Regarding the \textbf{machine-learning based retrieval} of communication episodes, all experts agreed that detecting related and sequential communication is \enquote{important for contextual information} (SI~1).
The visualization as density-based communication amount can be intuitively understood by all experts.
LEA~2 regarded it as beneficial that the detailed raw data from a communication episode can be inspected within the visualization.
Going on from there, the ability to train a machine learning model by giving communication episodes as examples is viewed as \enquote{opening up new possibilities} (RS~1).
Domain experts especially favored that \textbf{arbitrary features} can, in principle, be used for the machine learning model (extending upon those we defined above), which can be communication \enquote{based on text, audio, pictures, geographic information systems, or combined with graphs} (RS~1), using additional levels.
Therefore, the features itself \enquote{do not matter much, as the user has to define them}  (RS~1), making the flexibility of our approach \enquote{very broad} (RS~1).
With such broad applicability, the explainability and retracing of the results are 
\enquote{an important issue.
	If an analyst has a result, he needs to explain how he ended up there
} (LEA~3).
This explanation is simplified tremendously \enquote{with the generation of a step-by-step report} (LEA~3), as producible from our \textbf{provenance} history, whereas currently, \enquote{analysts have to write detailed accounts on how they got to the information} (LEA~3) and justify it each time in writing.
In terms of practical usage, they had \enquote{no ideas for additional [conceptual] features} (LEA~1), except for the inclusion of centrality measures.
But the potential possibilities with the framework are almost \enquote{overwhelming at first} (SI~1).
The \textbf{applicability} of the presented approach is not restricted to a narrow use case as the one presented.
Therefore, the system is 
\enquote{broadly applicable to multiple domains where you have bigger groups of communication data.
	Data that [law enforcement experts] often have to deal with.
	For example, organized crime, financial crime, or terrorism} (LEA~3).
All the other experts share this view.
Indeed, the presented system is \enquote{a beginning of an interaction platform where you can combine other logic's as well and [which] offers many possibilities} (RS~1), providing custom analysis levels specific for your needs (cf. RS~1).

\subsection{Case Study and User Evaluation}
\label{sec:evaluation_user_study}

During the formative expert assessment, the experts did not interact on their own with the system, as we were only able to secure a limited, non-individual amount of time with them.
To compensate for any possible bias, we additionally conducted a small-scale user study with two further domain experts (LEA~4 and RS~2).
For this we describe a possible case study which highlights the benefits of an holistic approach.

\vfill\eject

\textbf{Expertise}
LEA~4 is the head of the big data department of a federal governmental agency supporting criminal investigations.
She advises law enforcement agents on the applications of artificial intelligence to criminal investigations.
RS~2 is a senior scientist at a federal government research institute with over ten years of experience in communication analysis and terrorist investigations.

\textbf{Methodology and Case Study}
The case study is based on an artificially designed financial \textbf{fraud use-case} to identify senders and receivers of relevant communications.
The aim is to discover those persons which, during the first nine months of 2001, disseminated knowledge about legal issues involving persons in combination with organizations in California, and then identify the only person who received information from all of them.
The experts were given a system manual and a short written description of this task.
Such a task often occurs in real cases, but is not well supported by existing approaches using more advanced analysis concepts. Instead one often relies on keyword list using domain-specific ontologies which are difficult to create and require manual work to search through the results. Often, more advanced concepts are only presented in academic papers, but in isolation from each other and rarely combined in a single framework.
The experts were expected to map the task conditions to the analysis levels and their properties: namely specific time ranges, applicable field offices, and concepts that are relevant for the search (persons, organizations, and legal topics).
In a second step, when they have identified the persons that disseminated the knowledge, they were expected to make the mental transfer to restrict the view to those participants as sending persons and examine which participant is the only receiver.
The successful completion of the task was measured by checking the identified users.
During the task, the users were undisturbed to explore and try out the system freely.
It was the first time they saw and used the system for both domain experts, having no prior experience with it and having received no introduction.
After completion, we interviewed them shortly about their experience, which is described in the following.

\textbf{Findings}
Both experts were able to \textbf{successfully complete} the task within 15 minutes.
The interface and \textbf{interaction concepts} were described as \enquote{intuitive [and] self-explanatory} (LEA~4).
She was surprised to be offered a search for concepts and initially tried to use the keyword-based search level for conceptual searches.
After noticing no results and checking the \textbf{filter settings} again, she quickly realized her mistake and corrected it, using the visual query interface to build a conceptual query.
This leads us to conclude that some of the filters would benefit from a more detailed and hands-on explanation of their function.
The system offers \enquote{helpful} (RS~2) drill-downs of visualizations and is intuitive and straightforward (cf. RS~2) to use.
Both domain experts noted that the system naturally supports the investigative \textbf{workflows}, and the interaction design combined with the documentation is sufficient for working productively and getting relevant results.
Compared to their existing systems and workflows, the system provides a significant benefit in analytical capabilities.
Most notably, it allows the \textbf{simultaneous application} of different search methodologies to support cross-matches.
This results in a speed-up and allows for more powerful queries, in contrast to manually merging separate results.

\section{Discussion and Future Work} \label{sec:discussion}
During the formative expert assessment and the study, we received several proposals on how to extend our approach further.
Among \textbf{expected requests} for a research prototype like industry-grade interfaces, more supported data sets, and import features, are other more concerned with \textbf{core functionality} of the prototype.
To follow up on those, we discuss the limitations and the broader applicability of our approach in the following section and the context of future work.
For our prototype approach, we adapted the generic blueprint of modularized communication analysis to the case study by providing two example levels (not counting the standard task levels). \newline
Of course, the system can be extended modularly with \textbf{further analysis levels}, for example, those featuring graph centrality measures, community detection, leveraging specific meta-data like locations, or more specific content analysis modules based on linguistics.
Currently, this requires modification in the source code, but a plug-in architecture would be conceivable.
Regarding the overall \textbf{matrix visualization}, the challenges and limits of such an approach for communication analysis have partly been discussed in previous work~\cite{Fischer.HyperMatrix.2020}.
When tasks are primarily focused on network structure analysis, a classical node-link-diagram-based approach might be more suited.
Alternatively, one could consider a coordinated and linked view approach, for example, by adding a synced node-link-diagram as a separate main view of our approach.
\newline
Apart from adding new levels to this approach, a challenging step for future work is to investigate how this approach can be used for the analysis of \textbf{multi-party conversations}.
So far we duplicate messages with multiple recipients, which partly destroys the group aspect.
Future work could investigate how hypergraphs allow to capture such scenarios or how additional zoom level could be leveraged for group communication analysis.
\newline
Also, for the prototype development, we assumed some \textbf{practical restrictions}.
For instance, we require the receiver, sender, and timestamp information for each individual message.
This describes the structure of the network and forms the basis of the matrix-based visualization.
Another aspect of the prototype is the \textbf{visual query language}, where it could be valuable to extend the grammar and support nested queries visually.
\newline
Finally, the expert interviews and the user study resulted in a positive response to our workflow and prototype system.
However, we are well aware that the sample group for our formative experts' assessment and the user \textbf{study was quite limited}.
To achieve statistically more accurate results and broaden the perspective, the study could be extended with more participants as part of future work.
\newline
This explainable and accountable decision making is not only relevant in the security domain, in which the case study and expert assessment were conducted.
Indeed, the experts think the applications are not limited to such a narrow set of criminal communication investigations but can be applied to communication data in \textbf{other domains}.
One different application would be in the business intelligence domain.
The system could be applied as a search and retrieval mechanism to search for hidden, decentralized knowledge contained in business documents and communications.
This knowledge can then be linked and extracted into centralized knowledge management systems, allowing for more efficient management structures and avoiding redundancies, making the processes more accountable.

\section{Conclusion} \label{sec:conclusion}
So far, most interactive, automated communication analysis approaches focus either on the network aspects or on the content, in contradiction to communication research.
As such, the individual or isolated analysis may not suffice to capture the full available information and may lead to less effective, incomplete, and biased results.
Further, it can increase the struggle experts face when articulating their domain knowledge, not leveraging their full potential.
We address this challenge by arguing for and discussing a holistic approach to communication analysis, simultaneously applying both methods, allowing for more structured and detailed analytical capabilities.
To help domain experts deal with the complexity of modern communication data, we present \toolname{}, a blueprint for a visual analytics-based communication analysis system that offers a wider approach, providing a tight coupling between the network and the content analysis aspects, building on individual levels and supported by a machine learning-based retrieval system.
We provide two extendable levels as an example for network and content analysis each, covering dynamics modeling (based and extend from our previous work~\cite{SeebacherFischer.ConversationalDynamics.2019}) and semantic text analysis.
We leverage ideas from hypergraph analysis~\cite{Fischer.HyperMatrix.2020} for a multi-level matrix-based visualization design to integrate those levels in a single interface.
However, we specifically tailor and adapt this idea to communication analysis by providing specific visualization levels to support domain experts in their mental understanding during exploration and allow them to answer more detailed questions about communication behavior and structure, including identifying individual communication episodes.
Combining network and content aspects in a single visualization allows for maintaining overview and focus while eliminating demanding context switches, rapidly exploring large search spaces, and providing details on demand.
The realized techniques allow the simultaneous analysis of network and content aspects, like properties, conversational dynamics, or conceptual content, to refine the search and supports cross matches.
We evaluate our approach in one case study and through assessments with law enforcement experts using real-world communication data.
The results demonstrate that our system surpasses existing solutions, enabling the effective analysis of large amounts of information in a targeted and integrated way.
The experts regard this approach as a novel and promising way for a more meaningful communication analysis that can readily be applied to comprehensive analytical tasks as encountered in practical applications.
While we focused on communication analysis for law enforcement as driving application, many tasks in communication analysis are similar and, therefore, our methods are more generically applicable to a wider variety of domains, like digital humanities or business intelligence.
By bridging this gap between network and content analysis in automated communication analysis systems, we aim to pave the path for a more holistic approach to communication analysis.
\section*{Acknowledgments}
\noindent The authors acknowledge the financial support by the Federal Ministry of Education and Research of Germany (BMBF) in the framework of PEGASUS under the program "Forschung für die zivile Sicherheit 2018 - 2023" and its announcement "Zivile Sicherheit - Schutz vor organisierter Kriminalität II". \hfill \break \vspace*{-17pt}

\clearpage
\bibliographystyle{eg-alpha-doi}
\bibliography{bibliography}


\end{document}